\documentclass[%
 reprint,
 amsmath,amssymb,
 aps,
]{revtex4-2}

\usepackage{amssymb}
\usepackage{lipsum}
\usepackage{amsmath}

\usepackage{hyperref}
\hypersetup{
    colorlinks=true,
    linkcolor=red,
    filecolor=magenta,     
    citecolor=blue,
    urlcolor=magenta,
    pdfpagemode=FullScreen,
    breaklinks=false,
    }

\usepackage{graphicx}
\usepackage{dcolumn}
\usepackage{bm}

\begin{document}

\title{Optical Algorithm for Derivative of Real-Valued Functions}

\author{Murilo H. Magiotto}
\author{Guilherme L. Zanin}%
 \author{Wesley B. Cardoso} 

\author{Ardiley T. Avelar}

\author{Rafael M. Gomes}\email{rafaelgomes@ufg.br}
\affiliation{%
Instituto de Física, Universidade Federal de Goiás, Goiânia, 74690-900, GO, Brazil}

\date{\today}

\begin{abstract}
The derivation of a function is a fundamental tool for solving problems in calculus. Consequently, the motivations for investigating physical systems capable of performing this task are numerous. Furthermore, the potential to develop an optical computer to replace conventional computers has led us to create an optical algorithm and propose an experimental setup for implementing the derivative of one-dimensional real-valued functions using a paraxial and monochromatic laser beam. To complement the differentiation algorithm, we have experimentally implemented a novel optical algorithm that can transfer a two-dimensional phase-encoded function to the intensity profile of a light beam. Additionally, we demonstrate how to implement the n-th derivative of functions encoded in the phase of the transverse profile of photons.
\end{abstract}

\maketitle


\section{Introduction \label{I}}

New technologies, such as those associated with Artificial Intelligence, demand increasingly efficient computers due to the growing volume of data being processed, which overwhelms electronic processors. This trend suggests that the replacement of electronic computers, or their integration with another data processing platform, appears to be inevitable \cite{Greengard}. Optical computing has emerged as a promising alternative to electronic computing. Despite the challenges that remain for the large-scale production of computers with optical processors, their advantages over electronic computers offer hope for the development of a new and more efficient generation of computers. The processing of neural networks has recently spurred the search for practical models of optical computers \cite{Greengard, McMahon, Wetzstein, Shastri, Fahrenkopf, Tait}.

The main advantages of optical computing compared to conventional computers are related to low energy costs, bandwidth, spatial parallelism, among others \cite{Greengard, McMahon}. Another interesting aspect of optical computers is related to the possibility of integration with optical quantum computers \cite{Rudolph, 52325}. In this case, the advantage over electronic computers becomes even greater, since quantum physical effects can solve tasks that are impossible to perform with classical optical or electronic computing.  

The transverse profile of a paraxial and monochromatic light beam offers an alternative optical platform for implementing computational tasks such as real function integration \cite{Walborn}, integration of optical cryptographic systems and Deep Learning algorithms \cite{Qingming}, quantum computing \cite{52325,75201}, and quantum information \cite{62323, PhysRevA.72.022313, Laryssa}. In this study, we propose an algorithm using paraxial light beams capable of determining the magnitude of the n-th derivative of two-dimensional real-valued functions. We encode the function $f(x,y)$ into the phase of a linearly polarized monochromatic paraxial light beam using a Spatial Light Modulator (SLM). A spherical thin lens is employed to implement the Fourier transformation on the beam's transverse profile. In this context, we present the derivative algorithm using the photon's transverse degrees of freedom, with light polarization serving as the auxiliary variable for this task. The proposed scheme is applicable to light beams with bunching, random, and anti-bunching photons, making it suitable for both classical and quantum computing. Additionally, we propose and experimentally realize a novel method to transfer the phase-encoded function to the field intensity profile as a preliminary step in implementing the derivative algorithm. The experimental configurations described in this work can be readily implemented on silicon photonics platforms \cite{Tait}. Although the experimental setup can be easily adapted to the single-photon quantum regime by attenuating the laser beam \cite{Walborn06}, this adaptation does not provide a particular advantage in the final result of this specific algorithm.

This work is presented as follows: In Section \ref{I}, we introduce the work that will be presented here. In Section \ref{II}, we present the physical system that can be alternatively used to implement the optical algorithm. In Section \ref{III}, we will present an optical algorithm for the derivative of one-dimensional real-valued functions, accompanied by an experimental setup that can be used for practical implementation of the algorithm. In Section \ref{IV}, we propose and experimentally implement an optical algorithm capable of transferring the function $f(x,y)$ from the phase to the beam intensity profile. In Section \ref{V}, we extend the optical algorithm to the n-th partial derivative of a two-dimensional real-valued functions. In the last section (\ref{VI}), we present the conclusions of the results obtained.

\section{Transverse Continuous Variable \label{II}}

We propose an optical system to implement a derivation algorithm in the transverse profile of a laser beam. The complex amplitude of the field can be represented by
\begin{equation}
\textbf{U}(\textbf{r}) = E(\textbf{r})e^{-\imath kz}\boldsymbol{\epsilon}, 
\end{equation}
where $E(\textbf{r})$ is the complex envelope and can be identified as a spatial function of the transverse profile of the beam, $k = 2\pi /\lambda$ is a wavenumber and $\boldsymbol{\epsilon}$ is the polarization vector. An interesting feature arises when the complex envelope varies slowly in the z direction comparing the wavelength $\lambda$, as happens with laser beams. In this condition, the derivative in the variable z of the complex envelope is considered small, implying an even smaller 2nd derivative, which can be ignored. This is called the paraxial approximation, and the Helmholtz equation becomes:
\begin{equation}
    \frac{\imath}{k}\frac{\partial}{\partial z} E(\textbf{r}) = - \frac{1}{2k^2}\left( \frac{\partial^2}{\partial x^2} + \frac{\partial^2}{\partial y^2}  \right)E(\textbf{r}). \label{Eq. 2}
\end{equation}
This equation is known as the Helmholtz paraxial equation \cite{SALEH}. Due to the possibility of integrating the setup used here with a quantum computer, it is important to mention that Eq. \eqref{Eq. 2} is analogous to the Shcrödinger equation considering the variable z of the complex envelope as the time variable \cite{Stoler}.

\section{One-Dimensional Derivative \label{III}}

\begin{figure}[!h]
	\centering
		\includegraphics[scale=0.5]{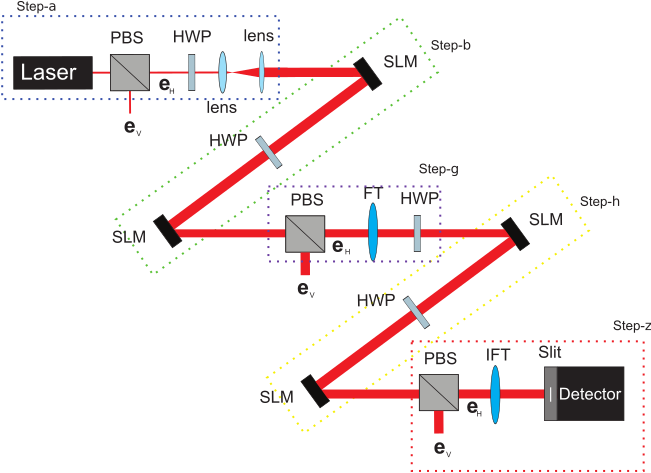}
	\caption{Proposed experimental setup.}
	\label{experiment2}
\end{figure}

In this section, we provide a detailed, step-by-step presentation of the algorithm and a proposed experimental setup for the practical implementation of the differentiation of real-valued functions by analyzing the intensity of a light beam after it traverses a system comprising lenses, polarizers, and spatial light modulators (see Fig. \ref{experiment2}). The experimental apparatus is described in detail through the complete schematic (Fig. \ref{experiment2}), which is segmented into sequential components labeled with indices ($\alpha, \beta, \gamma, \eta, \zeta$). Each of these components corresponds to a specific step in the optical algorithm.

\begin{figure}[tb]
	\centering
		\includegraphics[scale=0.80]{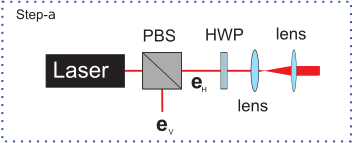}
	\caption{Step-$\alpha$: preparation of the input field.}
	\label{step1}
\end{figure}

In step-$\alpha$ (Fig. \ref{step1}), a laser beam is directed to a polarizing beam splitter (PBS) to verify the beam's polarization (which is horizontal in this case - $\textbf{e}_{H}$). If the polarization of the light beam is already linear and well-defined at the laser output, the initial PBS may be omitted. The beam then passes through a half-wave plate (HWP) with its fast axis oriented at $22.5^\circ$ relative to the horizontal direction (hereafter denoted as $HWP(22.5^\circ)$). After the HWP, two narrow lenses are used to increase the beam width.

To simplify the analysis, we will consider only one transverse dimension by omitting the $y$ variable in the representation of the complex envelope of the field. Additionally, we will designate the plane of the first SLM as the initial plane ($z=0$). Consequently, after step-$\alpha$, the resulting beam can be represented as follows
\begin{equation}
	\textbf{E}(x,z=0) =  \frac{E(x,0)}{\sqrt 2}(\textbf{e}_H + \textbf{e}_V),
\end{equation} 
where $\textbf{e}_H$ and $\textbf{e}_V$ denote the horizontal and vertical components of the field, respectively. Furthermore, we employ a set of lenses to expand the width of the light beam, thereby ensuring that the transverse profile of the field can be approximated as nearly constant within the effective area of the SLM and zero outside of it, i.e.,
\begin{equation}
E(x,0) =     \left\{ \begin{array}{rcl}
 E(x_0,0) & \mbox{if} & -x_{L} \leq x \leq x_{L} \\ 
 0  & \mbox{if} & x \leq -x_{L} \text{ or } x\geq x_{L},
 \end{array}\right.
\end{equation}
where 2$x_{L}$ is the horizontal lenght of the SLM. The position $x_{0}$ can be chosen between the interval $[-x_{L},x_{L}]$. Then, the initial field can be written as
\begin{equation}
	\textbf{E}(x,0) =  \frac{E(x_{0},0)}{\sqrt 2}(\textbf{e}_H + \textbf{e}_V),
\end{equation} 
where $E(x_{0},0)=const.$.

\begin{figure}[tb]
	\centering
	\includegraphics[scale=0.80]{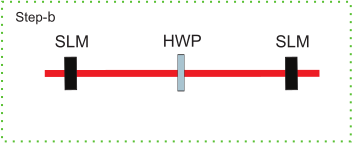}
	\caption{Step-$\beta$: printing of the function f(x) in the transverse profile of the beam.}
	\label{step2}
\end{figure}

In step-$\beta$, as depicted in Figure \ref{step2}, the beam is directed to the SLM to imprint the phase $e^{i a(x)}$, which is a function of the transverse coordinate $x$, onto the horizontal component of the beam. The SLM in this configuration modulates the phase in only one polarization state of the field; in this case, it is calibrated to modulate the phase in the horizontal polarization. The phase function $a(x)$ programmed into the SLM is derived from the function $f(x)$, which will be differentiated. The field immediately after passing through the SLM will be:
\begin{equation}
	\textbf{E}(x,0) =  \frac{E(x_{0},0)}{\sqrt 2}\left[e^{ i  a(x)}\textbf{e}_H + \textbf{e}_V\right].
\end{equation}
Note in Fig. \ref{step2} that a HWP will be used to rotate the polarization of the field again by $45^\circ$. In this way, the resulting field, slightly before the second SLM will be
\begin{eqnarray}
\textbf{E}(x,0) &=&  E(x_{0},0) e^{ i\frac{a(x)}{2}}\left[\cos\left(\frac{a(x)}{2}\right)\textbf{e}_H \right. \nonumber \\
 &+& \left. \sin\left(\frac{a(x)}{2}\right)\textbf{e}_V \right].
\end{eqnarray}
To $a(x) =\arccos(2f(x))$, the function above become,
\begin{equation}
	\textbf{E}(x,0) = E(x_{0},0)e^{ i  \frac{\arccos(2f(x))}{2}}\left[f(x)\textbf{e}_H +  i  g(x)\textbf{e}_V\right],
\end{equation}
with  $g(x) = \sin(\arccos(2f(x)))$.

To eliminate the global phase $e^{\imath \arccos(2f(x))/2}$, we can use a second SLM to imprint the phase $e^{ i b(x)}$, where $b(x) = -\arccos(2f(x))/2$. Then the field on the final of the step-$\beta$ is,
\begin{equation}
	\textbf{E}(x,0) = E(x_{0},0)\left[ f(x)\textbf{e}_H +  i h(x)\textbf{e}_V  \right],
\end{equation}
with $h(x) =  e^{ i  \arccos(2f(x))/2}g(x)$.

\begin{figure}[tb]
	\centering
	\includegraphics[scale=0.80]{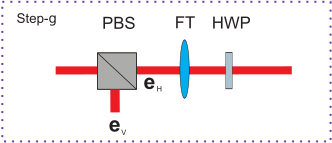}
	\caption{Step-$\gamma$: application of the Fourier transform. }
	\label{step3}
\end{figure}

In step-$\gamma$ (Fig. \ref{step3}), the beam is directed through a PBS to discard the $\textbf{e}_V$ component and subsequently passes through a HWP oriented at $22.5^\circ$. As a result, the field transforms to
\begin{equation}
	\textbf{E}(x,0) = E(x_{0},0)f(x)\left[\frac{\textbf{e}_H + \textbf{e}_V}{\sqrt 2 }\right].
\end{equation}

Next, a thin lens is used to implement the fourier transform ($\mathcal{F}$) of the transverse component of the field,
\begin{equation}
	\hat{\textbf{E}}(k) =  E(x_{0},0)\hat{f}(k)\left[\frac{\textbf{e}_H + \textbf{e}_V}{\sqrt 2 }\right].
\end{equation}
where $\hat{\textbf{E}}(k)=\mathcal{F}\{\textbf{E}(x,0)\}$ and $\hat{f}(k)=\mathcal{F}\{f(x)\}$.

\begin{figure}[tb]
	\centering
	\includegraphics[scale=0.80]{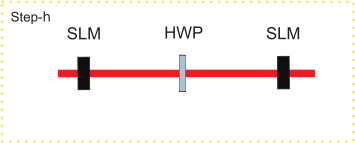}
	\caption{Step-$\eta$: printing of the $kf(k)$ function in the beam profile. }
	\label{step4}
\end{figure} 

Step-$\eta$: at this stage, the beam is again reflected into an SLM responsible for imprinting the phase $e^{ i c(k)}$ on the beam, now in coordinates of reciprocal space, resulting:
\begin{equation}
	\hat{\textbf{E}}(k) = E(x_{0},0)\hat{f}(k)\left[\frac{e^{ i  c(k)}\textbf{e}_H + \textbf{e}_V}{\sqrt 2 }\right],
\end{equation}	
where $ c(k) = arccos(2k)$.

In sequence, a $HWP(22.5^\circ)$ change the polarization of the beam that can be rewritten to
\begin{eqnarray}
	\hat{\textbf{E}}(k)	&=&   E(x_{0},0)\hat{f}(k)e^{ i arccos(2k)/2}[k\textbf{e}_H \nonumber \\ 
 &+&  i m(k)\textbf{e}_V],
\end{eqnarray}
where $m(k) = \sin(\arccos(2k))$.
The global phase $e^{ i arccos(2k)/2}$ can be eliminated using other SLM in sequence. In this way, the resulting beam is 
\begin{eqnarray}
\hat{\textbf{E}}(k) &=&   E(x_{0},0)\hat{f}(k)\left[ k \textbf{e}_H +  i n(k)\textbf{e}_V\right]\nonumber,\\
\end{eqnarray}
with $n(k) =  e^{ i   arccos(2k)/2}m(k)$.

\begin{figure}[tb]
	\centering
	\includegraphics[scale=0.80]{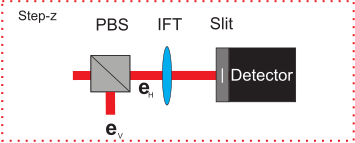}
	\caption{Step-$\zeta$: application of the inverse Fourier transform and detection. }
	\label{step5}
\end{figure}

For the last step of the experimental implementation of the optical algorithm (Fig. \ref{step5}), once again we eliminate the vertical component of the beam with $PBS$, so the resulting field is 
\begin{eqnarray}
	\hat{\textbf{E}}(k) &=&   E(x_{0},0)k\hat{f}(k) \textbf{e}_H. \label{Eq. 14}
\end{eqnarray}

In accordance with the property of the Fourier Transform $\mathcal{F}\{ f^{(n)}(x)\} = (- i k)^n\tilde{f}(k)$ \cite{RafWesArd}, the single lens show in the Fig. \ref{step5} is responsible to transform the field above (Eq. \eqref{Eq. 14}) into 
\begin{eqnarray}
	\textbf{E}(x,0) &=&   -i  E(x_{0},0)\frac{d}{dx}f(x)\hat{e}_H.
\end{eqnarray}

Finally, the beam must be detected point by point using a slit in the detection plane. The slit, centered at the point $x_0$ is responsible for printing the function $g(x-x_0)$ on the profile beam. In this way, the electric field becomes
\begin{eqnarray}
	\textbf{E}(x,0) &=&  - \imath g(x-x_0) {E}(x_0,0)\frac{d}{dx}f(x)\textbf{e}_H.
\end{eqnarray}

Then, the intensity of the beam will be
\begin{eqnarray}
	I &=& \int_{-x_l}^{x_l}|E(x,0)|^2 dx\nonumber\\
	 &=& |E(x_0,0)|^2 \int_{-x_l}^{x_l}\left|g(x - x_0) \frac{d}{dx}f(x)\right|^2dx\nonumber\\
	  &=& |E(x_0,0)|^2 \int_{-x_l}^{x_l}|g(x - x_0)|^2\left| \frac{d}{dx}f(x)\right|^2dx,
\end{eqnarray}
where 2$x_{l}$ is the width of the slit. Considering $g(x-x_0)$ as a Gaussian function of width $\sigma$, then its quadratic module is given by 
\begin{eqnarray}
|g(x-x_0)|^2 = \frac{1}{\sqrt{\pi}\sigma}e^{-(x-x_0)^2/\sigma^{2}} .
 \end{eqnarray}

In the Gaussian width limit tending to zero, the function $|g(x-x_0)|^2$ tends to a Dirac delta $\delta(x-x_0)$. Then, the beam intensity will be
\begin{eqnarray}
 	 I_{x_0} &=& |E(x_0,0)|^2 \lim_{\sigma \rightarrow 0} \int_{-x_l}^{x_l}|g(x - x_0)|^2\left| \frac{d}{dx}f(x)\right|^2dx\nonumber\\
 	  &=& |E(x_0,0)|^2 \int_{-x_l}^{x_l}\delta(x-x_0)\left| \frac{d}{dx}f(x)\right|^2dx\nonumber\\
 	   &=& |E(x_0,0)|^2 \left| \frac{d}{dx}f(x)\big|_{x=x_0} \right|^2 ,
 \end{eqnarray}
 i.e., the beam intensity is proportional to the derivative of the function $f(x)$ at the point $x_0$. It is important to note that, in this context, the intensity can be interpreted as the beam power \cite{SALEH}.

The therm $|E(x_0,0)|^2$ can be identified as the intensity (or power) of the beam when the SLM's not imprint phase ($I_i=|E(x_0,0)|^2$). As mentioned at the beginning of the text, the beam is sufficiently large to consider the transverse component of the field ($E(x,0)=E(x_0,0)$) as a constant at the plane $z=0$. Then, the normalized intensity of the beam in the $x_{0}$ is 
\begin{equation}
\frac{I_{x_0}}{I_{i}} = \left|\frac{d}{dx}f(x)\big|_{x=x_0}\right|^2
\end{equation}
then,
\begin{equation}
    \left.\frac{d}{dx}f(x)\right|_{x=x_0} = \pm\sqrt{\frac{I_{x_0}}{I_{i}}}.
\end{equation}
To complete the algorithm, it is necessary to determine the concavity of the function. Therefore, in the following section, we present a new optical algorithm designed to address this task.

\section{Concavity of the Function \label{IV}}

\begin{figure}[tb]
	\centering
	\includegraphics[scale=0.55]{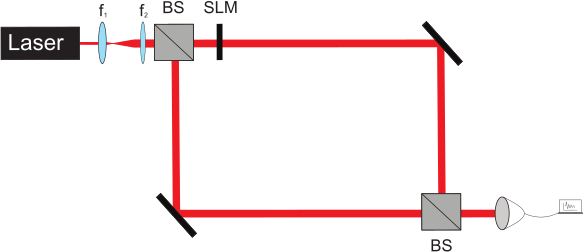}
	\caption{The experimental scheme for interfering two light beams is as follows: The initial two spherical lenses are used to expand the beam width. In one arm of the Mach-Zehnder interferometer, the phase modulation is implemented using a transmission SLM. In the other arm, a transmitting SLM is recommended to symmetrize the arms and mitigate any unintended phase effects. The intensity distribution can be analyzed using a CCD camera.}
	\label{Inter}
\end{figure}

To determine the concavity of the function $f(x,y)$, interference between two beams can be utilized. The intensity distribution resulting from the interference 
between the fields $E(x,y)e^{ia(x,y)}$ and $E(x,y)$ can be expressed by
\begin{equation}
I(x,y))=2|E(x,y)|^2[1+cos(a(x,y))].
\end{equation}
To obtain this intensity profile in practice, a Mach-Zehnder interferometer can be employed, incorporating a transmission modulator in one of the arms to introduce the phase difference $a(x,y)$ between the arms as shown in  Fig. \ref{Inter}. For $a(x,y)=arccos[f(x,y)]$, the resulting intensity will be
\begin{equation}
\frac{I(x,y)}{2I_{0}(x,y)}=1+f(x,y),
\end{equation}
where $I_{0}(x,y)=|E(x,y)|^2$. As mentioned above, the input field will be considered constant in the area of action of the modulators, we can take the term $I_{0}(x,y)$ as being constant $(I_{0})$, then 
\begin{equation}
f(x,y)=\frac{I(x,y)}{2I_{0}}+1.
\end{equation}

\begin{figure}[tb]
	\centering
	\includegraphics[scale=0.70]{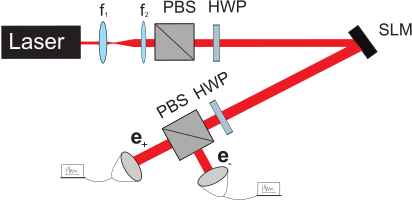}
	\caption{Experimental setup implemented to analyze the behavior of the function $f(x,y)$ through measurements on the field intensity distribution.}
	\label{Stokes}
\end{figure}

Another experimental approach to characterize the behavior of the function $f(x,y)$ involves measuring the Stokes parameter $S_{2}$ \cite{Walborn}. The experimental setup depicted in Fig. \ref{Stokes} is similar to the configuration used by Lemos et al. \cite{Walborn}, albeit with some significant modifications.

In Fig. \ref{Stokes}, the initial field $E(x)\hat{e}_H$ passing through the half-wave plate ($HWP22.5^\circ$) and reflecting in the SLM, becomes
\begin{equation}
   \hat{E}(x,y) =E(x,y)\frac{(e^{ia(x,y)}\hat{e}_H+\hat{e}_V)}{\sqrt{2}}.
\end{equation}
Rewriting the field in terms of the diagonal and antidiagonal polarization ($\hat{e}_{+}=(\hat{e}_{H}+\hat{e}_{V})/\sqrt{2}$ and $\hat{e}_{-}=(\hat{e}_{H}-\hat{e}_{V})/\sqrt{2}$), results
\begin{equation}
   \hat{E}(x,y) =E(x,y)\left(\frac{1+e^{ia(x,y)}}{2}\hat{e}_+ +\frac{1-e^{ia(x,y)}}{2}\hat{e}_-)\right).
\end{equation}
Finally, measuring the field intensity in diagonal ($I(x,y)^{+}$) and antidiagonal ($I(x,y)^{-}$) polarization using a PBS and a HWP (Fig. \ref{Stokes}), we obtain

\begin{equation}
I^{\pm}(x,y)=|E(x,y)|^{2}(1\pm cos[a(x,y)]).
\end{equation}
Note that the Stokes parameter $S_2(x,y)$ is obtained by $S_{2}(x,y)=I^{+}(x,y)-I^{-}(x,y)$. Thus, if $a(x,y)=arccos[f(x,y)]$, we obtain

\begin{equation}
f(x,y)=S_2(x,y)/I_0.
\end{equation}
In this way, using a CCD camera or scanning a point detector on both PBS outputs, we can obtain the function $f(x,y)$ through measurements of the intensity profiles $I^{+}$ and $I^{-}$. 

\subsection{Experiment}

To implement the experimental configuration shown in Fig. \ref{Stokes}, we utilized a 633 nm continuous-wave laser. The subsequent two lenses ($f_1 = 25$ mm and $f_2 = 75$ mm) serve to expand the beam width by a factor of 3. Following the lenses, the laser beam passes through a PBS and a HWP to prepare the field's polarization before reflecting onto a SLM (Holoeye Pluto NIR), where the phase $a(x,y)=arccos[f(x,y)]$ is encoded. The analysis of the intensity profile ($I^+$ and $I^{-}$) is conducted using an HWP and a PBS to select the polarizations $\hat{e}_{+}$ and $\hat{e}_{-}$ prior to detection by the camera.

\subsection{Results\label{IV-A}}

\begin{figure}[tb]
	\centering
	\includegraphics[scale=0.40]{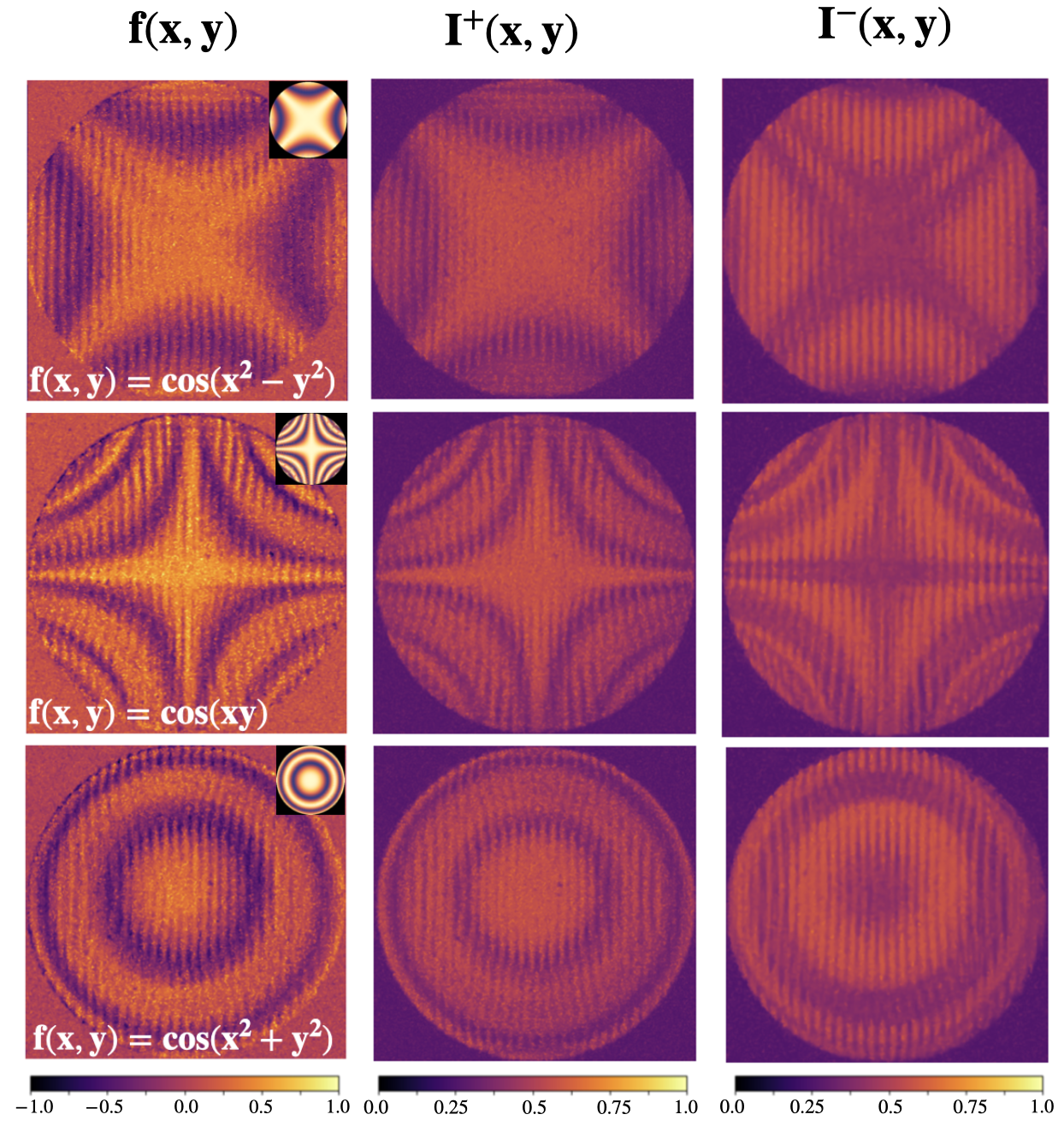}
	\caption{Experimental Results. For more details, see Sec. (\ref{IV-A})}.
	\label{Results}
\end{figure}

By implementing the experimental setup (Fig. \ref{Stokes}), we obtained the results shown in Fig. \ref{Results}. As indicated at the top of the figure, the first column represents the results obtained through the measurements of the intensity profile $I^{+}(x,y)$ (second column) and $I^{-}(x,y)$ (third column) for $f(x,y)=cos(x^2-y^2)$ (first row), $f(x,y)=cos(xy)$ (second row) and $f(x,y)=cos(x^2+y^2)$ (third row). The minifigures in the upper right corner of the subfigures in the first column represent the theoretical graph of the functions used in the experiment.

As we can see in Fig. \ref{Results}, the regions where the function is increasing and decreasing are evident. Note also that the experimental setup (Fig. \ref{Stokes}) can be obtained by adapting the setup shown in Fig. \ref{experiment2}. Thus, this step can be implemented before the derivation for calibration of the transverse plane, leaving known the regions where the derivative of the function will be positive and negative, thus completing the optical derivation algorithm.

\section{N-th partial derivative of two-dimensional functions. \label{V}}

The procedure for performing the n-th partial derivative of a two-dimensional real-valued function ($f(x,y)$) is analogous to the process described in the previous section, with minor changes.

Now, the two transverse components of the field will be used in the algorithm ($\textbf{E}(x,y,0)$). As was done in the previous section, the field in the plane of the first SLM ($z=0$) will be given by
\begin{eqnarray}
	\textbf{E}(x,y,0) =  \frac{E(x,y,0)}{\sqrt 2}(\textbf{e}_H + \textbf{e}_V).
\end{eqnarray} 
After being submitted to the $\beta$ step, the resulting field will be
\begin{eqnarray}
	\textbf{E}(x,y,0) &=& E(x_{0},y_{0},0)\left[ f(x,y)\textbf{e}_H +  i  g(x,y)\textbf{e}_V  \right]\nonumber,\\
\end{eqnarray}
where $g(x,y) = e^{ i  \arccos(2f(x,y))/2}\sin(\arccos(2f(x,y)))$ and $E(x_0,y_0,0)$ is defined as constant in the area of the SLM.

Then, after crossing the PBS and $HWP(22.5^\circ)$, the field undergoes the action of the lens that implements the Fourier Transform of the field ($\hat{\textbf{E}}(k_{x},k_{y})=\mathcal{F}\{\textbf{E}(x,y,0)\}$) in step-$\gamma$:
\begin{eqnarray}
	\hat{\textbf{E}}(k_{x},k_{y}) &=&E(x_{0},y_{0},0)\hat{f}(k_{x},k_{y})\left[\frac{\textbf{e}_H + \textbf{e}_V}{\sqrt 2 }\right].
\end{eqnarray}
where $\hat{f}(k_x,k_y)=\mathcal{F}\{f(x,y)\}$.

In the step-$\eta$, the phase $c(k_x, k_y) = arccos(2k_x^nk_y^m)$ is imprinted resulting in 
 \begin{equation}
\hat{\textbf{E}}(k_x,k_y) =   E(x_{0},y_{0})\hat{f}(k_x,k_y)\left[ k_x^nk_y^m \textbf{e}_H +  i  h(k_x,k_y)\textbf{e}_V\right],\nonumber\\
\end{equation}
where $h(k_x,k_y) = e^{ i   arccos(2k_x^nk_y^m)/2}\sin[\arccos(k_xk_y)]$.

In the final step, after the Inverse Fourier Transform, the field is 
\begin{equation}
	\textbf{E}(x,y,0) =   (-\imath)^{(n+m)}  E(x_0,y_0,0)\frac{\partial^n}{\partial x^n}\frac{\partial^m}{\partial y^m}f(x,y)\textbf{e}_H.
\end{equation}

To detect the field intensity a point detector is directed to the transverse position $(x,y)$ where the function will be derived. As in the case of one dimension, the measured intensity will be 
\begin{equation}
 I_{x_0,y_0}^{n,m}   = | E(x_{0},y_{0},0)|^2  \left| \frac{\partial^n}{\partial x^n}\frac{\partial^m}{\partial y^m}f(x,y) \big|_{x = x_0,y = y_0} \right|^2 .
 \end{equation}
Finally, the quadratic modulus of the n-th partial derivative of $f(x,y)$ is 

\begin{equation}
\left| \frac{\partial^n}{\partial x^n}\frac{\partial^m}{\partial y^m}f(x,y)\big|_{x = x_0,y = y_0} \right|^2  = \frac{I_{x_0,y_0^{n,m}}}{I_{in}},
\end{equation}
then, 
\begin{equation}
 \frac{\partial^n}{\partial x^n}\frac{\partial^m}{\partial y^m}f(x,y)\big|_{x = x_0,y = y_0}  =\pm\sqrt{\frac{I_{x_0,y_0}^{n,m}}{I_{in}}}.
\end{equation}
Where $I_{in}$ is the intensity (or power) of the beam when the SLM acts only as a mirror, i.e. $I_{in}=|E(x_{0},y_{0},0)|^2$. Additionally, the preliminary algorithm presented in the previous section (Sec. \ref{IV}) can be employed to characterize the regions where the function is increasing or decreasing. This approach effectively completes the optical algorithm for computing the n-th derivative of real-valued functions.

\section{Conclusion \label{VI}}

We demonstrate the feasibility of an optical algorithm designed for computing the derivative of one-dimensional real-valued functions. By utilizing spatial light modulators and the principles of the Optical Fourier Transform, we encode the function to be differentiated into the transverse profile of the paraxial and monochromatic light beam. By measuring the intensity of the beam, we can determine the derivative at a specified location determined by a slit positioned in front of the detector. Light polarization degrees of freedom are utilized as auxiliary variables within the presented algorithm. Additionally, we have experimentally implemented a novel optical algorithm for transferring the function encoded in the phase to the intensity profile of the light beam. Based on the measurements presented in Section \ref{IV}, we conclude that this supplementary optical algorithm effectively characterizes the regions of the measurement plane where the derivative is positive or negative, thus completing the differentiation process. Furthermore, our protocol can be readily extended to compute the n-th partial derivative of two-dimensional real functions. The simplicity and compatibility of this algorithm with existing platforms for optical computing (both classical and quantum) may foster further research into the use of the transverse profile of the field for information processing and mathematical operations.

\section*{Acknowledgements}
The authors acknowledge the financial support of the Brazilian agencies CNPq (Projeto Universal	No. 407469/2021-4, CNPq - PQ - 2022 No.	312173/2022-9, No. 306105/2022-5 and Sisphoton Laboratory-MCTI No. 440225/2021-3), CAPES, and FAPEG. This work was also performed as part of the Brazilian National Institute of Science and Technology (INCT) for Quantum Information (Grant No. 465469/2014-0).


\begin{thebibliography}{18}%
\makeatletter
\providecommand \@ifxundefined [1]{%
 \@ifx{#1\undefined}
}%
\providecommand \@ifnum [1]{%
 \ifnum #1\expandafter \@firstoftwo
 \else \expandafter \@secondoftwo
 \fi
}%
\providecommand \@ifx [1]{%
 \ifx #1\expandafter \@firstoftwo
 \else \expandafter \@secondoftwo
 \fi
}%
\providecommand \natexlab [1]{#1}%
\providecommand \enquote  [1]{``#1''}%
\providecommand \bibnamefont  [1]{#1}%
\providecommand \bibfnamefont [1]{#1}%
\providecommand \citenamefont [1]{#1}%
\providecommand \href@noop [0]{\@secondoftwo}%
\providecommand \href [0]{\begingroup \@sanitize@url \@href}%
\providecommand \@href[1]{\@@startlink{#1}\@@href}%
\providecommand \@@href[1]{\endgroup#1\@@endlink}%
\providecommand \@sanitize@url [0]{\catcode `\\12\catcode `\$12\catcode `\&12\catcode `\#12\catcode `\^12\catcode `\_12\catcode `\%12\relax}%
\providecommand \@@startlink[1]{}%
\providecommand \@@endlink[0]{}%
\providecommand \url  [0]{\begingroup\@sanitize@url \@url }%
\providecommand \@url [1]{\endgroup\@href {#1}{\urlprefix }}%
\providecommand \urlprefix  [0]{URL }%
\providecommand \Eprint [0]{\href }%
\providecommand \doibase [0]{https://doi.org/}%
\providecommand \selectlanguage [0]{\@gobble}%
\providecommand \bibinfo  [0]{\@secondoftwo}%
\providecommand \bibfield  [0]{\@secondoftwo}%
\providecommand \translation [1]{[#1]}%
\providecommand \BibitemOpen [0]{}%
\providecommand \bibitemStop [0]{}%
\providecommand \bibitemNoStop [0]{.\EOS\space}%
\providecommand \EOS [0]{\spacefactor3000\relax}%
\providecommand \BibitemShut  [1]{\csname bibitem#1\endcsname}%
\let\auto@bib@innerbib\@empty
\bibitem [{\citenamefont {Greengard}(2021)}]{Greengard}%
  \BibitemOpen
  \bibfield  {author} {\bibinfo {author} {\bibfnamefont {S.}~\bibnamefont {Greengard}},\ }\bibfield  {title} {\bibinfo {title} {Photonic processors light the way},\ }\href {https://doi.org/10.1145/3474357} {\bibfield  {journal} {\bibinfo  {journal} {Commun. ACM}\ }\textbf {\bibinfo {volume} {64}},\ \bibinfo {pages} {16–18} (\bibinfo {year} {2021})}\BibitemShut {NoStop}%
\bibitem [{\citenamefont {McMahon}(2023)}]{McMahon}%
  \BibitemOpen
  \bibfield  {author} {\bibinfo {author} {\bibfnamefont {P.~L.}\ \bibnamefont {McMahon}},\ }\bibfield  {title} {\bibinfo {title} {The physics of optical computing},\ }\href {https://doi.org/10.1038/s42254-023-00645-5} {\bibfield  {journal} {\bibinfo  {journal} {Nat. Rev. Phys.}\ }\textbf {\bibinfo {volume} {5}},\ \bibinfo {pages} {717} (\bibinfo {year} {2023})}\BibitemShut {NoStop}%
\bibitem [{\citenamefont {Wetzstein}\ \emph {et~al.}(2020)\citenamefont {Wetzstein}, \citenamefont {Ozcan}, \citenamefont {Gigan}, \citenamefont {Fan}, \citenamefont {Englund}, \citenamefont {Solja{\v c}i{\'c}}, \citenamefont {Denz}, \citenamefont {Miller},\ and\ \citenamefont {Psaltis}}]{Wetzstein}%
  \BibitemOpen
  \bibfield  {author} {\bibinfo {author} {\bibfnamefont {G.}~\bibnamefont {Wetzstein}}, \bibinfo {author} {\bibfnamefont {A.}~\bibnamefont {Ozcan}}, \bibinfo {author} {\bibfnamefont {S.}~\bibnamefont {Gigan}}, \bibinfo {author} {\bibfnamefont {S.}~\bibnamefont {Fan}}, \bibinfo {author} {\bibfnamefont {D.}~\bibnamefont {Englund}}, \bibinfo {author} {\bibfnamefont {M.}~\bibnamefont {Solja{\v c}i{\'c}}}, \bibinfo {author} {\bibfnamefont {C.}~\bibnamefont {Denz}}, \bibinfo {author} {\bibfnamefont {D.~A.~B.}\ \bibnamefont {Miller}},\ and\ \bibinfo {author} {\bibfnamefont {D.}~\bibnamefont {Psaltis}},\ }\bibfield  {title} {\bibinfo {title} {{Inference in artificial intelligence with deep optics and photonics}},\ }\href {https://doi.org/10.1038/s41586-020-2973-6} {\bibfield  {journal} {\bibinfo  {journal} {Nature}\ }\textbf {\bibinfo {volume} {588}},\ \bibinfo {pages} {39} (\bibinfo {year} {2020})}\BibitemShut {NoStop}%
\bibitem [{\citenamefont {Shastri}\ \emph {et~al.}(2021)\citenamefont {Shastri}, \citenamefont {Tait}, \citenamefont {Ferreira~de Lima}, \citenamefont {Pernice}, \citenamefont {Bhaskaran}, \citenamefont {Wright},\ and\ \citenamefont {Prucnal}}]{Shastri}%
  \BibitemOpen
  \bibfield  {author} {\bibinfo {author} {\bibfnamefont {B.~J.}\ \bibnamefont {Shastri}}, \bibinfo {author} {\bibfnamefont {A.~N.}\ \bibnamefont {Tait}}, \bibinfo {author} {\bibfnamefont {T.}~\bibnamefont {Ferreira~de Lima}}, \bibinfo {author} {\bibfnamefont {W.~H.~P.}\ \bibnamefont {Pernice}}, \bibinfo {author} {\bibfnamefont {H.}~\bibnamefont {Bhaskaran}}, \bibinfo {author} {\bibfnamefont {C.~D.}\ \bibnamefont {Wright}},\ and\ \bibinfo {author} {\bibfnamefont {P.~R.}\ \bibnamefont {Prucnal}},\ }\bibfield  {title} {\bibinfo {title} {{Photonics for artificial intelligence and neuromorphic computing}},\ }\href {https://doi.org/10.1038/s41566-020-00754-y} {\bibfield  {journal} {\bibinfo  {journal} {Nat. Photonics}\ }\textbf {\bibinfo {volume} {15}},\ \bibinfo {pages} {102} (\bibinfo {year} {2021})}\BibitemShut {NoStop}%
\bibitem [{\citenamefont {Fahrenkopf}\ \emph {et~al.}(2019)\citenamefont {Fahrenkopf}, \citenamefont {McDonough}, \citenamefont {Leake}, \citenamefont {Su}, \citenamefont {Timurdogan},\ and\ \citenamefont {Coolbaugh}}]{Fahrenkopf}%
  \BibitemOpen
  \bibfield  {author} {\bibinfo {author} {\bibfnamefont {N.~M.}\ \bibnamefont {Fahrenkopf}}, \bibinfo {author} {\bibfnamefont {C.}~\bibnamefont {McDonough}}, \bibinfo {author} {\bibfnamefont {G.~L.}\ \bibnamefont {Leake}}, \bibinfo {author} {\bibfnamefont {Z.}~\bibnamefont {Su}}, \bibinfo {author} {\bibfnamefont {E.}~\bibnamefont {Timurdogan}},\ and\ \bibinfo {author} {\bibfnamefont {D.~D.}\ \bibnamefont {Coolbaugh}},\ }\bibfield  {title} {\bibinfo {title} {{The AIM Photonics MPW: A Highly Accessible Cutting Edge Technology for Rapid Prototyping of Photonic Integrated Circuits}},\ }\href {https://doi.org/10.1109/JSTQE.2019.2935698} {\bibfield  {journal} {\bibinfo  {journal} {IEEE J. Sel. Top. Quantum Electron.}\ }\textbf {\bibinfo {volume} {25}},\ \bibinfo {pages} {1} (\bibinfo {year} {2019})}\BibitemShut {NoStop}%
\bibitem [{\citenamefont {Tait}\ \emph {et~al.}(2019)\citenamefont {Tait}, \citenamefont {Ferreira~de Lima}, \citenamefont {Nahmias}, \citenamefont {Miller}, \citenamefont {Peng}, \citenamefont {Shastri},\ and\ \citenamefont {Prucnal}}]{Tait}%
  \BibitemOpen
  \bibfield  {author} {\bibinfo {author} {\bibfnamefont {A.~N.}\ \bibnamefont {Tait}}, \bibinfo {author} {\bibfnamefont {T.}~\bibnamefont {Ferreira~de Lima}}, \bibinfo {author} {\bibfnamefont {M.~A.}\ \bibnamefont {Nahmias}}, \bibinfo {author} {\bibfnamefont {H.~B.}\ \bibnamefont {Miller}}, \bibinfo {author} {\bibfnamefont {H.-T.}\ \bibnamefont {Peng}}, \bibinfo {author} {\bibfnamefont {B.~J.}\ \bibnamefont {Shastri}},\ and\ \bibinfo {author} {\bibfnamefont {P.~R.}\ \bibnamefont {Prucnal}},\ }\bibfield  {title} {\bibinfo {title} {{Silicon Photonic Modulator Neuron}},\ }\href {https://doi.org/10.1103/PhysRevApplied.11.064043} {\bibfield  {journal} {\bibinfo  {journal} {Phys. Rev. Appl.}\ }\textbf {\bibinfo {volume} {11}},\ \bibinfo {pages} {064043} (\bibinfo {year} {2019})}\BibitemShut {NoStop}%
\bibitem [{\citenamefont {Rudolph}(2017)}]{Rudolph}%
  \BibitemOpen
  \bibfield  {author} {\bibinfo {author} {\bibfnamefont {T.}~\bibnamefont {Rudolph}},\ }\bibfield  {title} {\bibinfo {title} {{{Why I am optimistic about the silicon-photonic route to quantum computing}}},\ }\href {https://doi.org/10.1063/1.4976737} {\bibfield  {journal} {\bibinfo  {journal} {APL Photonics}\ }\textbf {\bibinfo {volume} {2}},\ \bibinfo {pages} {030901} (\bibinfo {year} {2017})}\BibitemShut {NoStop}%
\bibitem [{\citenamefont {Wallman}\ and\ \citenamefont {Emerson}(2016)}]{52325}%
  \BibitemOpen
  \bibfield  {author} {\bibinfo {author} {\bibfnamefont {J.~J.}\ \bibnamefont {Wallman}}\ and\ \bibinfo {author} {\bibfnamefont {J.}~\bibnamefont {Emerson}},\ }\bibfield  {title} {\bibinfo {title} {Noise tailoring for scalable quantum computation via randomized compiling},\ }\href {https://doi.org/10.1103/PhysRevA.94.052325} {\bibfield  {journal} {\bibinfo  {journal} {Phys. Rev. A}\ }\textbf {\bibinfo {volume} {94}},\ \bibinfo {pages} {052325} (\bibinfo {year} {2016})}\BibitemShut {NoStop}%
\bibitem [{\citenamefont {Lemos}\ \emph {et~al.}(2014)\citenamefont {Lemos}, \citenamefont {Ribeiro},\ and\ \citenamefont {Walborn}}]{Walborn}%
  \BibitemOpen
  \bibfield  {author} {\bibinfo {author} {\bibfnamefont {G.~B.}\ \bibnamefont {Lemos}}, \bibinfo {author} {\bibfnamefont {P.~H.~S.}\ \bibnamefont {Ribeiro}},\ and\ \bibinfo {author} {\bibfnamefont {S.~P.}\ \bibnamefont {Walborn}},\ }\bibfield  {title} {\bibinfo {title} {{Optical integration of a real-valued function by measurement of a Stokes parameter}},\ }\href {https://doi.org/10.1364/JOSAA.31.000704} {\bibfield  {journal} {\bibinfo  {journal} {J. Opt. Soc. Am. A}\ }\textbf {\bibinfo {volume} {31}},\ \bibinfo {pages} {704} (\bibinfo {year} {2014})}\BibitemShut {NoStop}%
\bibitem [{\citenamefont {Zhou}\ \emph {et~al.}(2023)\citenamefont {Zhou}, \citenamefont {Zhang}, \citenamefont {Wang}, \citenamefont {Xu}, \citenamefont {Xue},\ and\ \citenamefont {Zhang}}]{Qingming}%
  \BibitemOpen
  \bibfield  {author} {\bibinfo {author} {\bibfnamefont {Q.}~\bibnamefont {Zhou}}, \bibinfo {author} {\bibfnamefont {L.}~\bibnamefont {Zhang}}, \bibinfo {author} {\bibfnamefont {X.}~\bibnamefont {Wang}}, \bibinfo {author} {\bibfnamefont {B.}~\bibnamefont {Xu}}, \bibinfo {author} {\bibfnamefont {J.}~\bibnamefont {Xue}},\ and\ \bibinfo {author} {\bibfnamefont {Y.}~\bibnamefont {Zhang}},\ }\bibfield  {title} {\bibinfo {title} {Optical encryption using a sparse-data-driven framework},\ }\href {https://doi.org/https://doi.org/10.1016/j.optlaseng.2023.107825} {\bibfield  {journal} {\bibinfo  {journal} {Opt. Lasers Eng.}\ }\textbf {\bibinfo {volume} {171}},\ \bibinfo {pages} {107825} (\bibinfo {year} {2023})}\BibitemShut {NoStop}%
\bibitem [{\citenamefont {Sawada}\ and\ \citenamefont {Walborn}(2018)}]{75201}%
  \BibitemOpen
  \bibfield  {author} {\bibinfo {author} {\bibfnamefont {K.}~\bibnamefont {Sawada}}\ and\ \bibinfo {author} {\bibfnamefont {S.~P.}\ \bibnamefont {Walborn}},\ }\bibfield  {title} {\bibinfo {title} {{Experimental quantum information processing with the Talbot effect}},\ }\href {https://doi.org/10.1088/2040-8986/aac5c1} {\bibfield  {journal} {\bibinfo  {journal} {J. Opt.}\ }\textbf {\bibinfo {volume} {20}},\ \bibinfo {pages} {075201} (\bibinfo {year} {2018})}\BibitemShut {NoStop}%
\bibitem [{\citenamefont {Jiang}\ \emph {et~al.}(2007)\citenamefont {Jiang}, \citenamefont {Taylor}, \citenamefont {S\o{}rensen},\ and\ \citenamefont {Lukin}}]{62323}%
  \BibitemOpen
  \bibfield  {author} {\bibinfo {author} {\bibfnamefont {L.}~\bibnamefont {Jiang}}, \bibinfo {author} {\bibfnamefont {J.~M.}\ \bibnamefont {Taylor}}, \bibinfo {author} {\bibfnamefont {A.~S.}\ \bibnamefont {S\o{}rensen}},\ and\ \bibinfo {author} {\bibfnamefont {M.~D.}\ \bibnamefont {Lukin}},\ }\bibfield  {title} {\bibinfo {title} {Distributed quantum computation based on small quantum registers},\ }\href {https://doi.org/10.1103/PhysRevA.76.062323} {\bibfield  {journal} {\bibinfo  {journal} {Phys. Rev. A}\ }\textbf {\bibinfo {volume} {76}},\ \bibinfo {pages} {062323} (\bibinfo {year} {2007})}\BibitemShut {NoStop}%
\bibitem [{\citenamefont {Almeida}\ \emph {et~al.}(2005)\citenamefont {Almeida}, \citenamefont {Walborn},\ and\ \citenamefont {Souto~Ribeiro}}]{PhysRevA.72.022313}%
  \BibitemOpen
  \bibfield  {author} {\bibinfo {author} {\bibfnamefont {M.~P.}\ \bibnamefont {Almeida}}, \bibinfo {author} {\bibfnamefont {S.~P.}\ \bibnamefont {Walborn}},\ and\ \bibinfo {author} {\bibfnamefont {P.~H.}\ \bibnamefont {Souto~Ribeiro}},\ }\bibfield  {title} {\bibinfo {title} {Experimental investigation of quantum key distribution with position and momentum of photon pairs},\ }\href {https://doi.org/10.1103/PhysRevA.72.022313} {\bibfield  {journal} {\bibinfo  {journal} {Phys. Rev. A}\ }\textbf {\bibinfo {volume} {72}},\ \bibinfo {pages} {022313} (\bibinfo {year} {2005})}\BibitemShut {NoStop}%
\bibitem [{\citenamefont {Ferro}\ \emph {et~al.}(2024)\citenamefont {Ferro}, \citenamefont {Avelar}, \citenamefont {Cardoso},\ and\ \citenamefont {Gomes}}]{Laryssa}%
  \BibitemOpen
  \bibfield  {author} {\bibinfo {author} {\bibfnamefont {L.~F.}\ \bibnamefont {Ferro}}, \bibinfo {author} {\bibfnamefont {A.~T.}\ \bibnamefont {Avelar}}, \bibinfo {author} {\bibfnamefont {W.~B.}\ \bibnamefont {Cardoso}},\ and\ \bibinfo {author} {\bibfnamefont {R.~M.}\ \bibnamefont {Gomes}},\ }\bibfield  {title} {\bibinfo {title} {Single photon hybrid quantum key distribution},\ }\href {https://doi.org/10.1088/1402-4896/ad185c} {\bibfield  {journal} {\bibinfo  {journal} {Phys. Scr.}\ }\textbf {\bibinfo {volume} {99}},\ \bibinfo {pages} {025102} (\bibinfo {year} {2024})}\BibitemShut {NoStop}%
\bibitem [{\citenamefont {Walborn}\ \emph {et~al.}(2006)\citenamefont {Walborn}, \citenamefont {Lemelle}, \citenamefont {Almeida},\ and\ \citenamefont {Ribeiro}}]{Walborn06}%
  \BibitemOpen
  \bibfield  {author} {\bibinfo {author} {\bibfnamefont {S.~P.}\ \bibnamefont {Walborn}}, \bibinfo {author} {\bibfnamefont {D.~S.}\ \bibnamefont {Lemelle}}, \bibinfo {author} {\bibfnamefont {M.~P.}\ \bibnamefont {Almeida}},\ and\ \bibinfo {author} {\bibfnamefont {P.~H.~S.}\ \bibnamefont {Ribeiro}},\ }\bibfield  {title} {\bibinfo {title} {{Quantum Key Distribution with Higher-Order Alphabets Using Spatially Encoded Qudits}},\ }\href {https://doi.org/10.1103/PhysRevLett.96.090501} {\bibfield  {journal} {\bibinfo  {journal} {Phys. Rev. Lett.}\ }\textbf {\bibinfo {volume} {96}},\ \bibinfo {pages} {090501} (\bibinfo {year} {2006})}\BibitemShut {NoStop}%
\bibitem [{\citenamefont {Saleh}\ and\ \citenamefont {Teich}(1991)}]{SALEH}%
  \BibitemOpen
  \bibfield  {author} {\bibinfo {author} {\bibfnamefont {B.~E.~A.}\ \bibnamefont {Saleh}}\ and\ \bibinfo {author} {\bibfnamefont {M.~C.}\ \bibnamefont {Teich}},\ }\href {https://doi.org/https://doi.org/10.1002/0471213748.fmatter_indsub} {\emph {\bibinfo {title} {{Fundamentals of Photonics}}}}\ (\bibinfo  {publisher} {John Wiley \& Sons, Ltd},\ \bibinfo {year} {1991})\BibitemShut {NoStop}%
\bibitem [{\citenamefont {Stoler}(1981)}]{Stoler}%
  \BibitemOpen
  \bibfield  {author} {\bibinfo {author} {\bibfnamefont {D.}~\bibnamefont {Stoler}},\ }\bibfield  {title} {\bibinfo {title} {Operator methods in physical optics},\ }\href {https://doi.org/10.1364/JOSA.71.000334} {\bibfield  {journal} {\bibinfo  {journal} {J. Opt. Soc. Am.}\ }\textbf {\bibinfo {volume} {71}},\ \bibinfo {pages} {334} (\bibinfo {year} {1981})}\BibitemShut {NoStop}%
\bibitem [{\citenamefont {Gomes}\ \emph {et~al.}(2021)\citenamefont {Gomes}, \citenamefont {Cardoso},\ and\ \citenamefont {Avelar}}]{RafWesArd}%
  \BibitemOpen
  \bibfield  {author} {\bibinfo {author} {\bibfnamefont {R.~M.}\ \bibnamefont {Gomes}}, \bibinfo {author} {\bibfnamefont {W.~B.}\ \bibnamefont {Cardoso}},\ and\ \bibinfo {author} {\bibfnamefont {A.~T.}\ \bibnamefont {Avelar}},\ }\bibfield  {title} {\bibinfo {title} {{Proposal for Anderson localization in transverse spatial degrees of freedom of photons}},\ }\href {https://doi.org/https://doi.org/10.1016/j.optcom.2021.127225} {\bibfield  {journal} {\bibinfo  {journal} {Opt. Commun.}\ }\textbf {\bibinfo {volume} {498}},\ \bibinfo {pages} {127225} (\bibinfo {year} {2021})}\BibitemShut {NoStop}%
\end{thebibliography}
\end{document}